\documentclass{article}
\usepackage{graphics}
\usepackage{color}
\usepackage{epsfig}

\def \pt   {p_{\rm\scriptscriptstyle T}}
\newcommand\hatpt{\hat{p}_{\rm \scriptscriptstyle T}}
\newcommand\ep{\epsilon}

\begin{document}
\title{
\begin{flushright}
 \vspace*{-1.5in}
\begin{minipage}{3cm}
 {\normalsize Bicocca-FT-02-17\\[-2ex]
 July 2002} \\
\end{minipage}
\end{flushright}
\vspace*{0.5in}
HEAVY FLAVOUR PRODUCTION\footnote{Invited talk at the 6-th International
Workshop on Heavy Quarks and Leptons,
May 27/June 1, 2002, Vietri s/m, Salerno, Italy.}}
\author{
P. Nason \\
{\em INFN, Sez. di Milano, Milan, Italy}
}
\maketitle
\baselineskip=11.6pt
\begin{abstract}
Recent progress in Heavy Flavour Production phenomenology are discussed.
In particular,
the long-standing discrepancy of the Tevatron $b$ production data
is considered. It is shown that a better use of $e^+e^-$ data in constraining
the effect of fragmentation reduces the discrepancy considerably. 
\end{abstract}
\baselineskip=14pt
\section{Introduction}
The phenomenology of heavy flavour production has been so far a mixed success.
Although the order of magnitude of the total cross sections, and the shape
of differential distribution is reasonably predicted, large discrepancy are
present, especially for $b$ production. 

The theoretical framework for the description of heavy flavour production
is the QCD improved parton model. Besides the well-established
NLO corrections to the
inclusive production of heavy quark in hadron-hadron \cite{Nason:1988xz,
Nason:1989zy,Beenakker:1991ma,Mangano:1992jk},
hadron-photon \cite{Ellis:1989sb,Smith:1992pw,Frixione:1994dg},
and photon-photon collisions \cite{Drees:1993eh},
much theoretical work has been done
in the resummation of contributions enhanced in certain regions of phase
space: the Sudakov region, the large transverse momentum region
and the small-$x$ region.

The theoretical effort involved is justified by the large variety of
applications that heavy quark production physics has, in top,
bottom and charm production. 
Besides the need of modeling these processes, heavy quark production
is an important benchmark for testing QCD and parton model ideas,
due to the relative complexity of the production process,
the large range of masses available, and the existence of
different production environments, like $e^+e^-$ annihilation,
hadron, photon-hadron and photon-photon collisions.
\section{Total cross section for top and bottom}
\begin{figure}[htb]
\begin{center}
\epsfig{file=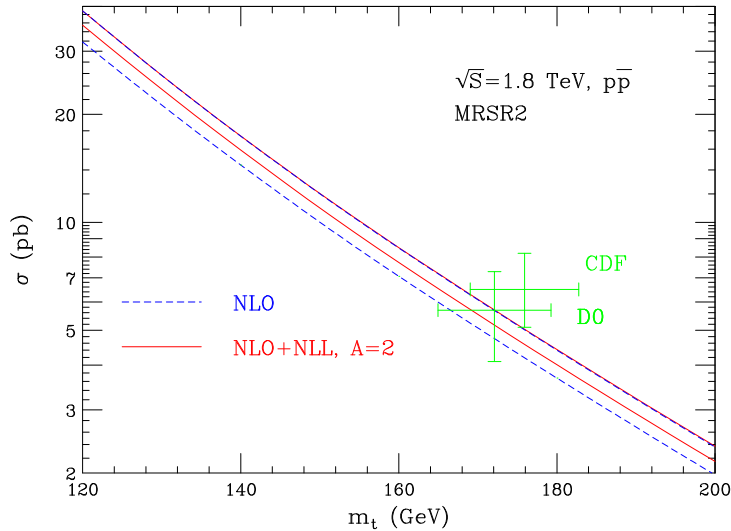,width=0.8\textwidth}
\end{center}
 \caption{\it
    Results on top cross sections at the Tevatron.
    \label{topxsec} }
\end{figure}

Top production \cite{Affolder:2001wd,Abazov:2002gy}
has been a most remarkable success of the theoretical
model. The measured cross section has been found in good agreement
with theoretical calculations, as shown in fig.~\ref{topxsec}.
Resummation of soft gluon effects \cite{Bonciani:1998vc} reduces the
theoretical uncertainty in the cross section, pushing it toward the
high side of the theoretical band. It remains, however, inside
the theoretical band of the fixed order calculation, thus showing
consistency with the estimated error.

Recently, the HERA-B experiment has measured the $b\bar{b}$ total
cross section \cite{Abt:2002rd}. This experiment is sensitive to the
moderate transverse momentum region, where the bulk of the total cross
section is concentrated. Since the production
is (in a certain sense) close to threshold, resummation of Sudakov effects
is useful also in this case.
\begin{figure}[htb]
\begin{center}
\epsfig{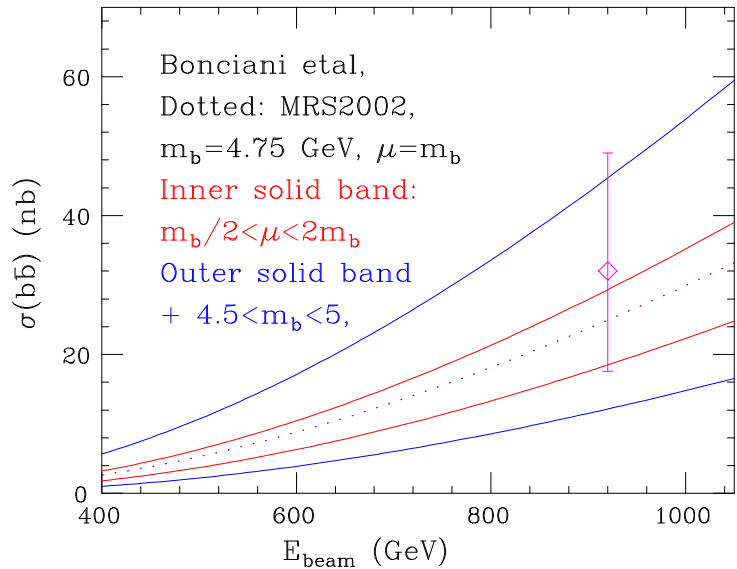}
\end{center}
 \caption{\it
    $b\bar{b}$ total cross section versus theoretical predictions,
    including resummation of soft gluons.
    \label{bxsec} }
\end{figure}
A comparison of the HERA-b measurement with a theoretical calculation
is shown in fig.~\ref{bxsec}.
The HERA-b result is compatible with the central value prediction,
with the $b$ pole mass around
$4.75$ GeV. Higher precision may constrain further the $b$ quark mass.
\section{Differential distributions}
CDF has a longstanding disagreement with QCD in $b$ production.
A very recent publication of the $B^+$ differential cross section
\cite{Acosta:2001rz} has quantified
the disagreement as a factor of $2.9 \pm 0.2 \pm 0.4$ in the ratio
of the measured cross section over the theoretical prediction.
This discrepancy has been present since a long time,
and it has been observed both in CDF and D0.
Some authors \cite{Berger:2000mp} have argued that the discrepancy could be
interpreted as a signal for Supersimmetry.

Because of the large theoretical uncertainties, this discrepancy has been often
downplayed. In fact, several effects may conspire to raise the $b$ cross
section to an appropriate value. It was early recognized that small-$x$
effects may be important in $b$ production at the Tevatron \cite{Nason:1988xz}.
Resummation of these effects \cite{Catani:1991eg,Collins:1991ty}
leads to an increase of the
cross section of the order of 30\%. Threshold effects are small in this
case \cite{Bonciani:1998vc}, being below 15\%, but they are
nevertheless positive.
Resummation of large $\log \pt$ yields an increase in the $\pt$ spectrum
of the order of 20\%\ in the intermediate $\pt$ region \cite{Cacciari:1998it}.
The presence of many possible enhancements should not lead, however,
to excessive optimism. First of all, it is not clear whether these effect
can be added up without overcounting. Furthermore, they are all higher
order effects, and thus should not push the cross section too far out
of the theoretical band, which includes estimates of unknown higher
order effects.

\begin{figure}[htb]
\begin{center}
\epsfig{file=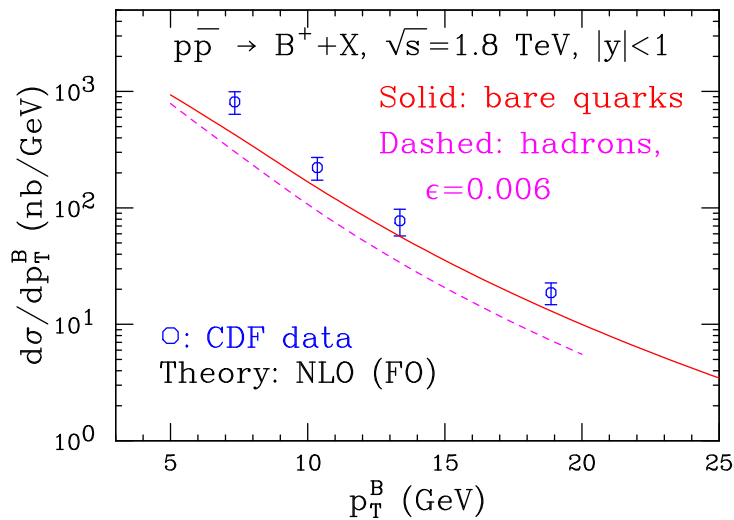,width=0.8\textwidth}
\end{center}
 \caption{\it
    Quark versus Hadron differential cross section compared to CDF data.
    The Hadron cross section is obtained by convoluting the quark cross
    section with a Peterson fragmentation function, with $\ep=0.006$.
    \label{bBdiff} }
\end{figure}

\begin{figure}[htb]
\begin{center}
\epsfig{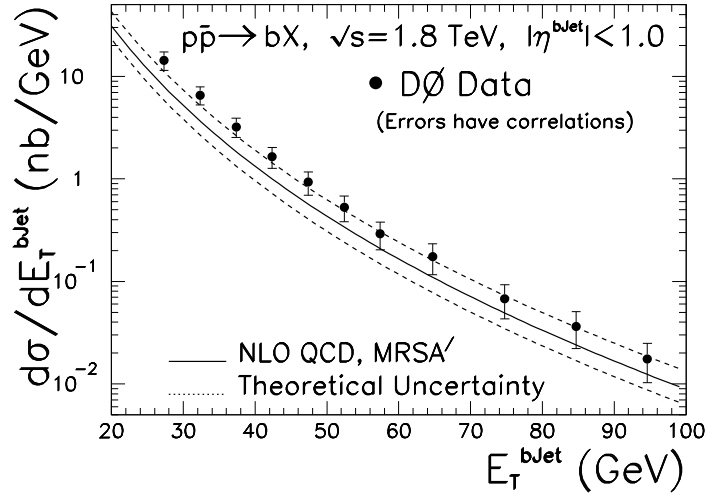}
\end{center}
 \caption{\it
    D0 $b$-jet data compared with the calculation of
    ref.~\cite{Frixione:1997nh}.
    \label{bjets} }
\end{figure}

It has been observed since some time that an improper understanding
of fragmentation effects may be one of the causes of the Tevatron
discrepancy. This possibility stems from the fact that $b$ quark
cross sections are in reasonable agreement with the Tevatron measurements
of the $B$ meson spectrum, while the cross section obtained by applying
a standard fragmentation function of the Peterson form \cite{Peterson:1983ak}
with $\ep=0.006$ to the quark cross section yields too soft a spectrum,
as can be seen in fig.~\ref{bBdiff}.
In ref.~\cite{Frixione:1997nh} it was suggested to study
$b$ quark jets rather than $B$ meson's distributions.
In fact, the jet momentum should be less sensitive to fragmentation
effects than the hadron momentum. A D0 study \cite{Abbott:2000iv} has
demonstrated that by considering $b$ jets instead of $B$ hadrons
the agreement between
theory and data improves considerably (see fig.~\ref{bjets}).
\section{Fragmentation}
If fragmentation effects are so important, they should be
assessed very carefully before claiming a discrepancy with data.
Non-perturbative fragmentation is usually introduced by
writing the hadron-level cross section for $B$ mesons as
\begin{equation}
\frac{d\sigma^B}{d\pt}
= \int d\hatpt dz \frac{d\sigma^b}{d\hatpt}
D(z) \;\delta(\pt-z\hatpt)\;,
\label{Bhad}
\end{equation}
where $\pt$ is the hadron, and $\hatpt$ is the quark transverse momentum.
Traditionally, $D(z)$ is given by the Peterson form \cite{Peterson:1983ak}
\begin{equation}\label{eq:peterson}
D(z)\propto\frac{z(1-z)^2}{((1-z)^2+\epsilon z)^2}\;.
\end{equation}
Strictly speaking, this procedure is justifiable only in the limit
of large transverse momenta (i.e. $p_T\gg m_b$). It can however be consider
a rough model of the effect of fragmentation at moderate and small
transverse momenta, as long as the variable $z$ is referred to the
momentum or the kinetic energy,
rather than the energy or the $+$ component, of the quark momentum.

Eq.~(\ref{eq:peterson}) is often used with $\ep=0.006$ for bottom
and $\ep=0.06$ for charm, in conjunction with shower Montecarlo programs.
These values of $\ep$ were obtained from fits to inclusive
heavy flavour production distributions in $e^+e^-$ collisions.
The shower Montecarlo accounts for hard fragmentation effects
(i.e. hard gluon radiation), where hard means from the typical
transverse momentum of the process down to some cutoff, of the
order of the hadron mass. This accounts for scaling violation
in the inclusive cross section for the production
heavy quarks\footnote{The inclusive cross
section for single particle production in $e^+e^-$ collisions
is sometimes also called fragmentation function, which is often a source
of confusion}.

It has become common practice to
adopt the value $\ep=0.006$ also in conjunction with NLO calculations
of $b$ production cross sections. This practice is not totally
justified, since the procedure used to obtain this value of $\ep$ does
not match the accuracy of the NLO calculation. A better approach
is possible; one can use NLL resummation of transverse momentum logarithms
in the context of $e^+e^-$ data in order to extract an appropriate
non-perturbative fragmentation function, and supplement NLO heavy
flavour production calculations with NLL resummation, in order to
correctly account for scaling violation. The theoretical tools
to do this are in fact available in the literature:
\begin{itemize}
\item[1] Single inclusive particle production in hadronic collisions
\cite{Aversa:1990uv}. Single hadron
production are described in term of NLO single parton cross section
convoluted with a NLL fragmentation function;
\item[2] Heavy quark Fragmentation Function \cite{Mele:1991cw};
a method for the computation of the heavy quark fragmentation function at
all orders in perturbation theory is developed, and applied at NLL.
Several applications in $e^+e^-$ physics have appeared
\cite{Colangelo:1992kh,Cacciari:1997wr,Cacciari:1997du,Nason:1999zj}.
\item[3] Single inclusive heavy quark production at large $p_T$
\cite{Cacciari:1994mq}; item 1 and 2 are combined to give
a NLL resummation of transverse momentum logarithms in heavy quark production;
\item[4] FONLL calculation of single inclusive heavy quark production;
item 3 is merged without overcounting with standard NLO
calculations. This procedure has been implemented both in
hadroproduction \cite{Cacciari:1998it} and
in photoproduction \cite{Cacciari:2001td,Frixione:2002zv}.
\end{itemize}
When using these ingredients, it was found that the CDF discrepancy
is considerably reduced \cite{Cacciari:2002pa}. It can be quantified as
a factor of
$1.7\pm 0.5\;\mbox{(theory)} \pm 0.5\; \mbox{(expt)}$.
\begin{figure}[htb]
\begin{center}
\epsfig{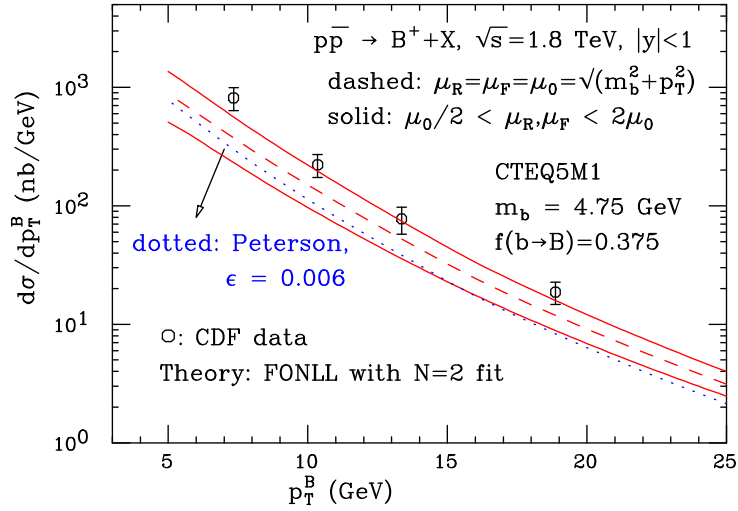}
\end{center}
 \caption{\it
Comparison of CDF data with the calculation of ref.~\cite{Cacciari:2002pa}.
 }
\label{Bhadcn}
\end{figure}
\begin{figure}[htb]
\begin{center}
\epsfig{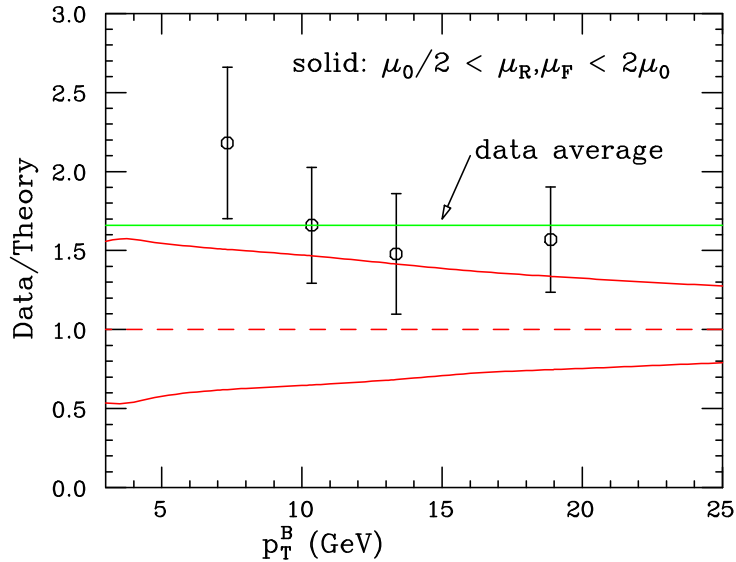}
\end{center}
 \caption{\it
As in fig.~\ref{Bhadcn}, ratios of data to theory.
\label{Bhadratcn}
 }
\end{figure}
Reduction in the discrepancy is due to four basic points:
\begin{itemize}
\item[1]
FONLL calculation brings about a 20\% increase
in the intermediate $p_T$ region.
\item[2]
The fragmentation step for going from a ``perturbative'' $b$ quark
to a $B$ hadron seems to be too strong at small $p_T$ in the
CDF implementation. We get a 20\% increase
in the small $p_T$ region.
\item[3]
A Peterson fragmentation function with $\epsilon=0.006$
is too soft (present $e^+e^-$ data favour values around $0.002$);
this is a 20\%\ effect.
\item[4]
A more accurate use of $e^+e^-$ input data on the fragmentation
function, described in the following section,
brings about another 20\% effect.
\end{itemize}
The net effect is $1.2^3=1.7$ and $2.9/1.7=1.7$,
so that the 2.9 CDF discrepancy becomes our $1.7$.
\section{A better use of $e^+e^-$ data on $b$ fragmentation}
$D(z)$ is extracted from fits to $e^+e^-$ data. One has
\begin{equation}
\underbrace{\frac{d\sigma(e^+e^-\to B+X)}{dp}}_{\mbox{measured}}=
\int d\hat{p}\, dz\;
\underbrace{\frac{d\sigma(e^+e^-\to b+X)}{d\hat{p}}}_{\mbox{computed}}
 \;\underbrace{D(z)}_{\mbox{fitted}} \;\delta(p-z\hat{p})
\end{equation}
From the above equation it is easy to show that
\begin{equation}
\langle x_B^{N-1} \rangle = \langle x_b^{N-1} \rangle \;D_N
\end{equation}
where
\begin{equation}
 x_B=\frac{p}{p_{\rm max}},\quad\quad
 x_b=\frac{\hat{p}}{\hat{p}_{\rm max}},\quad\quad
 D_N=\int dz z^{N-1} D(z).
\end{equation}
It turns out that for the computation of the hadronic
cross sections, only the first few
moments of the fragmentation function are important. This follows
from the fact that the heavy quark hadroproduction cross section
is a steeply falling function of the transverse momentum, following
roughly a power law. Assuming the form
\begin{equation}
\frac{ d\sigma(H_1 H_2\to b+X)}{d\hatpt} \approx A\hatpt^{-n}
\end{equation}
one finds
\begin{equation}\label{sigapprox}
\frac{d\sigma(H_1 H_2\to b+X)}{d\pt} 
\approx\int dz d\hatpt\, D(z) \frac{A}{\hatpt^n}
 \,\delta(\pt-z \hatpt) = \frac{A}{\pt^n} D_n=\frac{d\sigma}{d\hatpt}D_n\;.
\end{equation}
\noindent
This simple fact was noticed long ago
\cite{Frixione:1998ma,Nason:1999ta}.  The exponent $n$ ranges from 3
to 5 in hadronic collisions.  In ref.~\cite{Nason:1999ta} it was shown
that eq.~(\ref{sigapprox}) is also an excellent approximation to the
exact convolution.

\begin{figure}[htb]
\begin{center}
\epsfig{file=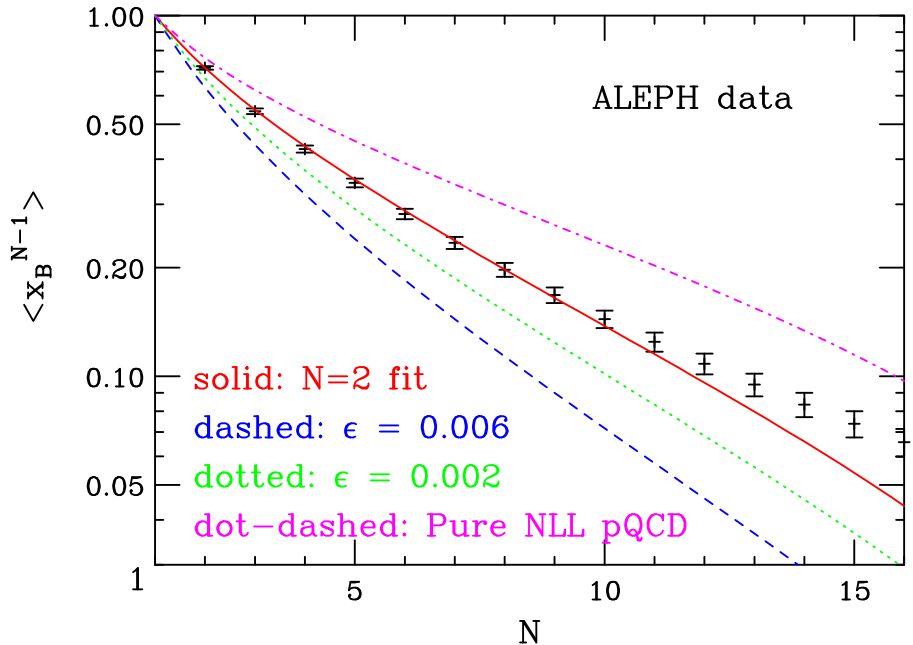,width=\textwidth}
\end{center}
 \caption{
The measured $\langle x_B^{N-1}\rangle$, 
compared with the perturbative
NLL calculation supplemented with different non-perturbative 
fragmentation forms.
The solid line is obtained using a one-parameter form fitted to the
second moment.
    \label{cfrkart} }
\end{figure}

When fitting the non-perturbative fragmentation function from $e^+e^-$ data,
the low moments relevant for hadroproduction are not well
fitted. The Peterson form with $\ep=0.002$
gives a good fit to the shape of the inclusive cross section, but not to the
moments. For the purpose of predicting hadronic cross section it is instead
better to make sure that the first few moments are accurately fitted.
In fig.~\ref{cfrkart}, it is shown that a fit to the second moment
alone with the simple form  $x(1-x)^\beta$, yields a good fit to all moments
in the interesting region (i.e. between 3 and 5).
We also notice that the second moment fit
($N=2$ fit from now on), the $\ep=0.002$ and the $\ep=0.006$ fits differ by
 20\% around the $N=4$ moment.

The reason why $x$ fits do not yield good fits to moments is easy to explain.
The large $x$ region in the inclusive production of heavy
flavoured hadrons in $e^+e^-$ annihilation is a very difficult region
to model. It is in fact affected by large perturbative (i.e.
Sudakov) and non-perturbative effects. It is unlikely that the simple
forms used commonly for the non-perturbative fragmentation function
can accurately model these complex effects.
Thus, the large $x$ region is usually excluded from
the fits. On the other hand, this region is the most important one for
moments.

From fig.~\ref{cfrkart}, another important fact becomes apparent.
When using NLL calculations for the perturbative part of the
fragmentation function, the importance of the non-perturbative
effects on moments is greatly reduced. This can be taken as an indication
of the fact that the heavy quark fragmentation function is,
to a considerable extent, perturbatively calculable.

\section{Other $B$ production data}
\begin{figure}[htb]
\begin{center}
\epsfig{file=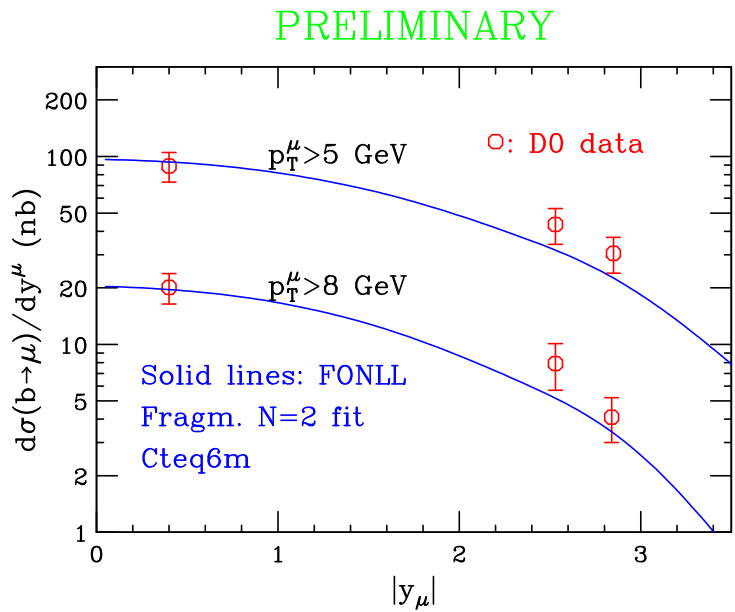,width=0.8\textwidth}
\end{center}
 \caption{\it \label{d0-fmu}
    D0 forward muon data compared with a calculation performed using the method
    of ref.~\cite{Cacciari:2002pa}.
    \label{D0fmu}}
\end{figure}

The procedure of ref.~\cite{Cacciari:2002pa}
may affect other $b$ production data
at hadron colliders. Here I consider, as an example, the inclusive muon cross
section in the central and forward region, where the muons come from
the semileptonic decays of the $b$ flavoured hadrons. These distributions
were measured by the D0\cite{Abbott:1999wu} experiment,
and found to be higher than theoretical prediction
by up to a factor of 4 in the forward region. The result of the
calculation performed with the method of ref.~\cite{Cacciari:2002pa}
is reported in fig.~\ref{d0-fmu}.
As one can see, the agreement with data is now quite remarkable.
\section{Conclusions}
The theory of heavy flavour production seems to give a good
qualitative description of the available data.
In the case of top production, the comparison between theory and experiment
is satisfactory also at a quantitative level.

Recent progress has taken place in the field of $b$ hadroproduction.
The HERA-b experiment has provided a cross section for $b$ production
at relatively low CM energy. Some progress in understanding the role
of fragmentation has considerably reduced the longstanding problem
of the $b$ momentum spectrum at the Tevatron.

Major problems do remain in the
(perhaps less developed) areas of bottom production in $\gamma\gamma$
and $\gamma p$ collisions. A discussion of these problems is given
in ref. \cite{Nason:2001kz}.
\providecommand{\href}[2]{#2}\begingroup\raggedright\endgroup

\end{document}